\documentclass[twocolumn,floatfix]{aastex63}
\usepackage{algorithm}
\usepackage{verbatim}
\usepackage{hyperref}
\usepackage{xspace} 
\xspaceaddexceptions{]\}}

\begin{document}

\title{Explaining ultra-massive quiescent galaxies at $3 < z < 5$ in the context of their environments}
\correspondingauthor{Christian Kragh Jespersen}
\email{ckragh@princeton.edu}

\author[0000-0002-8896-6496]{Christian Kragh Jespersen}
\affiliation{Department of Astrophysical Sciences, Princeton University, Princeton, NJ 08544, USA}

\author[0000-0002-1482-5818]{Adam C. Carnall}
\affiliation{Institute for Astronomy, University of Edinburgh, Royal Observatory, Edinburgh, EH9 3HJ, UK}

\author[0000-0001-7964-5933]{Christopher C. Lovell}
\affiliation{Institute of Cosmology and Gravitation, University of Portsmouth, Burnaby Road, Portsmouth, PO1 3FX, UK}

\begin{abstract}

The swift assembly of the earliest galaxies poses a significant challenge to our understanding of galaxy formation. Ultra-massive quiescent galaxies at intermediate redshifts ($3 < z < 5$) currently present one of the most pressing problems for theoretical modeling, since very few mechanisms can be invoked to explain how such galaxies formed so early in the history of the Universe. Here, we exploit the fact that these galaxies all reside within significant overdensities to explain their masses. To this end, we construct and release a modified version of the Extreme Value Statistics (EVS) code which takes into account galaxy environment by incorporating clustering in the calculation. With this new version of EVS, we find that ultra-massive quiescent galaxies at $3<z<5$ do not present as serious a tension with simple models of galaxy formation when the analysis of a given galaxy is conditioned on its environment.
\end{abstract}


\section{Introduction}
\label{sec:intro}

The James Webb Space Telescope is currently pushing the limits of galaxy formation theory. The discoveries of not only ultra-high redshift galaxies, but also extremely massive quiescent galaxies at $3 < z < 5$, imply that a surprisingly large amount of star formation took place at very early times \citep[e.g.,][]{Carnall2023_CEERS_Quiescent_Gals, Carnall2023_nature_GS9209,Valentino23_quiescentAtlas, Carniani2024_JADES-GS-z14-0, Glazebrook2024_highmass_quiescent_galaxy, deGraaf_highz_massive_quiescent}. Many possible explanations have been invoked \textit{post-hoc}, including highly bursty star formation, changes to $\Lambda$CDM, and artificially elevated star-formation efficiencies \citep{Dekel2023_bursts, Lovell2023_EVS, Endsley_2024_bursty_highz}. 

While bursty star formation seems like a promising explanation for observations of ultra-high redshift galaxies \citep{Mason2023_MUV_scatter}, it is irrelevant for inferring the masses of quiescent galaxies. Quiescent galaxies therefore present a unique opportunity for isolating other effects. This has resulted in suggestions that star formation efficiencies (SFE) must either be close to 100\%, or that modifications to the well-understood physics of halo formation would be necessary \citep{Carnall2024_EXCELS, Glazebrook2024_highmass_quiescent_galaxy}.

However, previous analyses of early ultra-massive quiescent galaxies have so far ignored the well-known fact that low redshift and simulated massive quiescent galaxies reside in overdense environments \citep{Frenk1988_halo_LCDM, Norberg2002_2CdF_clustering}, a relation which seems to continue even at high redshift, as massive quiescent galaxies observed with \textit{JWST} appear to preferentially populate overdensities \citep{Valentino23_quiescentAtlas, Jin2024_cosmic_vine, Ito2024_size_stellar_mass, Morishita2025_JWST_clusters}.

For example, the broadly used Extreme Value Statistics (EVS) approach of \cite{Lovell2023_EVS} assumes that all galaxies exist in a perfectly average field with no clustering. Another approach, presented by \cite{Jespersen2025_CV_highz}, marginalizes over the environment, since the environments of ultra-high redshift galaxies are rarely known. However, at intermediate redshifts, we may obtain strong constraints on the type of overdensity in which ultra-massive quiescent galaxies reside \citep{Overzier2016_protocluster_review}. These overdense environments require corrections to EVS \citep{davis_most_2011}. Here, we therefore explicitly include the dependence on overdensity in the EVS calculation of the probability distribution for the mass of the most massive galaxy observed in a given survey volume. We apply our model to three ultra-massive quiescent galaxies discovered with \textit{JWST} in the CANDELS UDS field.

\section{Data}
\label{sec:data}

\begin{figure*}
    \centering
    \includegraphics[trim={18.5cm 6cm 1cm 30cm},clip, width=0.98\linewidth]{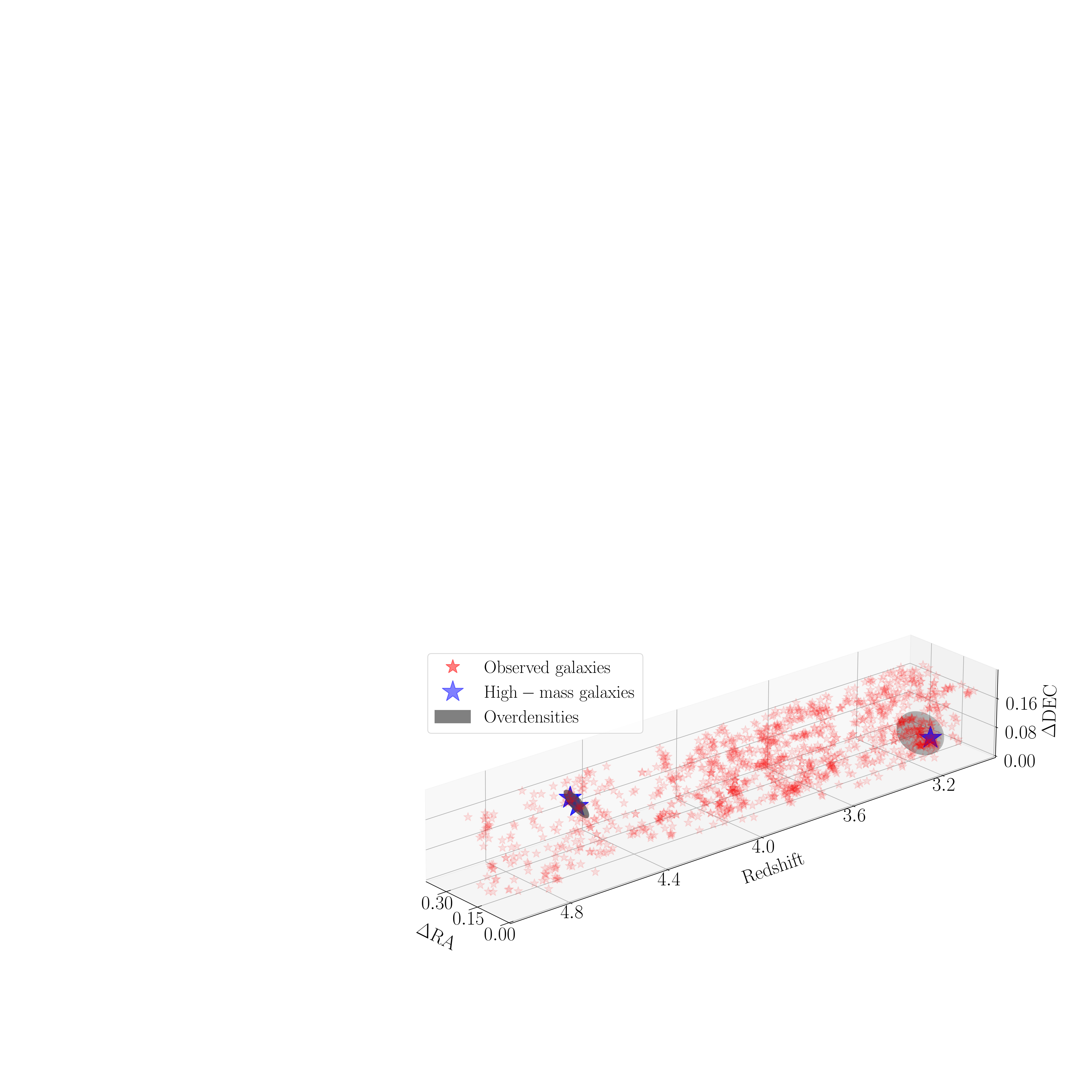}
    \begin{minipage}{0.475\linewidth}
        \centering
        \includegraphics[trim={9.5cm 4.3cm 1cm 13.0cm},clip, width=\linewidth]{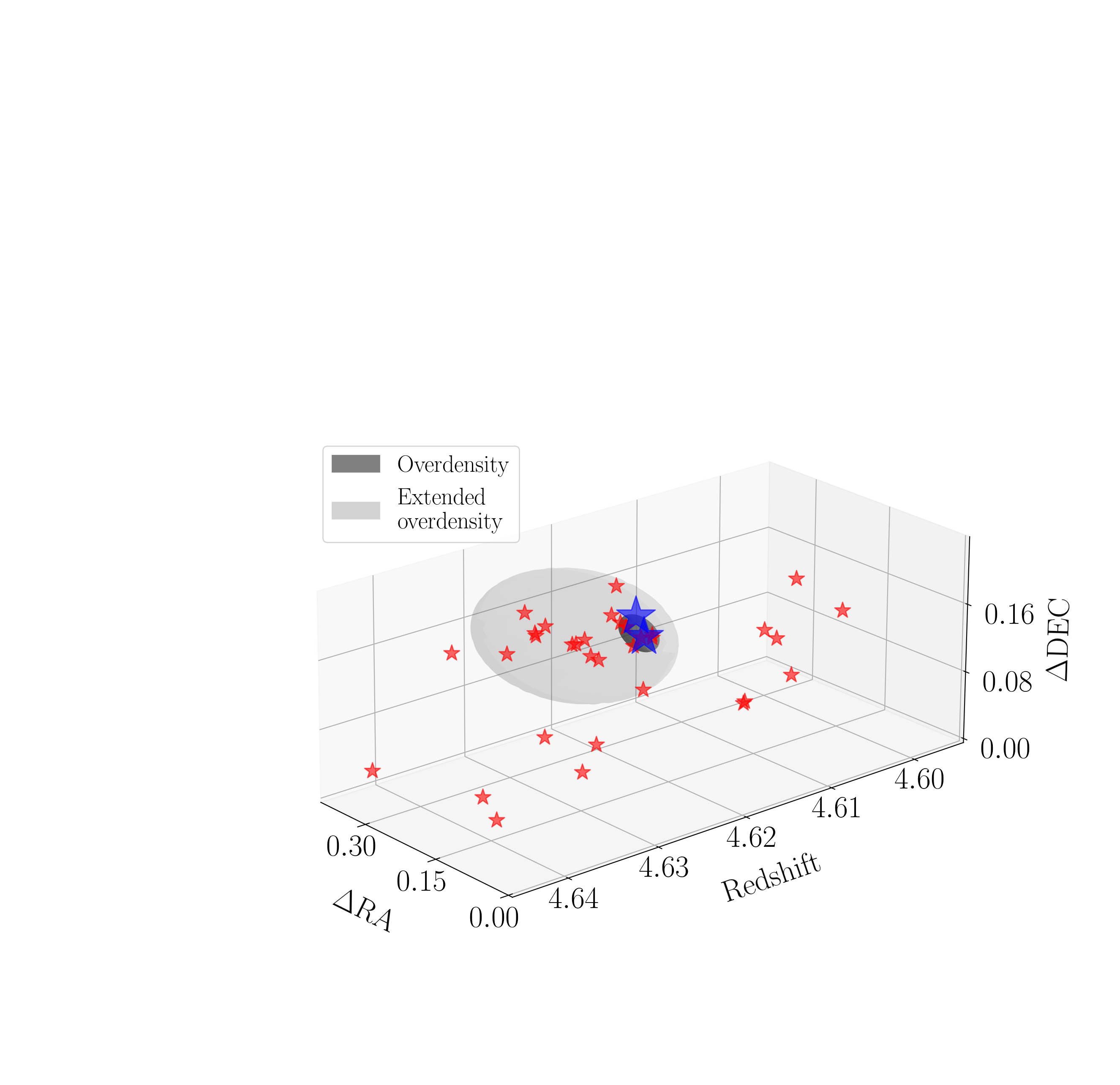}
    \end{minipage}
    \hfill
    \begin{minipage}{0.475\linewidth}
        \centering
        \includegraphics[trim={9.5cm 4.3cm 1cm 13.0cm},clip, width=\linewidth]{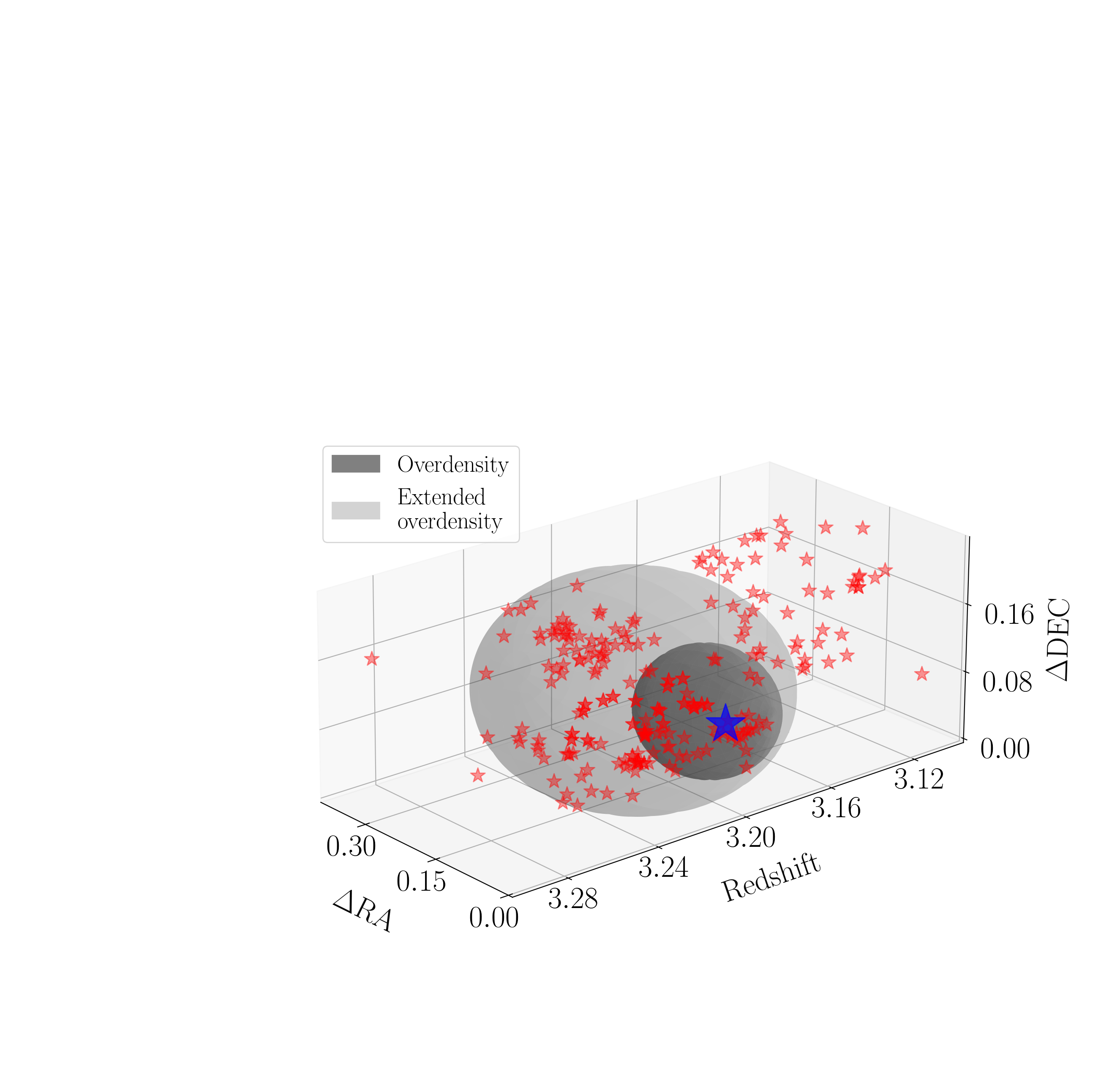}
    \end{minipage}
    \caption{\textbf{Top:} The 3D distribution of galaxies at $3 < z < 5$ with spectroscopic redshifts within the PRIMER UDS area. Strong clustering is visible, with two prominent overdensities (marked in grey). The overdensities host ultra-massive galaxies (blue stars), which have been proposed to challenge galaxy formation models. 
    \textbf{Bottom:} Further identification of possible concentrated and extended overdensities. 
    \textbf{Left:} Two possible $z\approx4.62$ volumes that could define the overdensity that PRIMER-EXCELS-109760/117560 exist in. The volumes are roughly 1.5 orders of magnitude apart.
    \textbf{Right:} Two possible $z\approx3.21$ volumes that could define the overdensity that ZF-UDS-7329 exists in. The volumes are roughly an order of magnitude apart. }
    \label{fig:EXCELS_overdensity}
\end{figure*}

In this work, we consider the objects discussed in \cite{Carnall2024_EXCELS} in the UDS field. They discuss three especially problematic galaxies: ZF-UDS-7329 at $z$=3.195 with log$_{10}(M_*/\mathrm{M_\odot})\simeq11.1$ (\citealt{Schreiber2018, Nanayakkara2024, Glazebrook2024_highmass_quiescent_galaxy}), PRIMER-EXCELS-109760 ($z=4.622$), and PRIMER-EXCELS-117560 ($z=4.621$), both with log$_{10}(M_*/\mathrm{M_\odot})\simeq11$ and newly reported by \cite{Carnall2024_EXCELS}. The relatively involved EXCELS selections are detailed in \S 3 of \cite{Carnall2024_EXCELS}, with \S 3.2.1 being of particular importance to the selection of the main galaxies of interest here.

To investigate the environments of these objects, we compile a list of available spectroscopic redshifts in the UDS field within the PRIMER survey area from $3 < z < 5$ as follows. We begin with the EXCELS spectroscopic redshift catalogue presented in \cite{Carnall2024_EXCELS}. We then supplement this with the DJA NIRSpec graded v3 compilation \citep{Heintz2024,deGraaff2025}, using only redshifts with a grade of 3. We supplement these with the compilation of pre-\textit{JWST} spectroscopic redshifts from \cite{Begley2025}.

The resulting distribution of objects with reliable spectroscopic redshifts in the PRIMER UDS field from $3 < z < 5$ is shown in Figure \ref{fig:EXCELS_overdensity}, with the three objects discussed in \cite{Carnall2024_EXCELS} highlighted in blue.

\section{Methods}
\label{sec:methods}

To calculate the mass of the most massive galaxy conditioned on density, a few steps must be taken. We quickly summarize these steps, and expand on them in the following subsections.

\begin{enumerate}
    \item The volumes of the overdensities must be identified (see \S \ref{subsec:overdensity_ID}).
    \item A mapping from the total possible number of such volumes in a given survey ($\mathrm{N_{\delta}}$) to a distribution of overdensities must be established. This is most easily done by estimating the \textit{density percentile} ($\mathrm{u_\delta}$) of a given overdensity (see \S \ref{subsec:sigma_delta_max}).
    \item Stellar mass functions dependent on density percentile, SMF($\mathrm{u_\delta}$), must be constructed (see \S \ref{subsec:model} and Figure \ref{fig:demo_fig}). 
\end{enumerate}

For each SMF($\mathrm{u_\delta}$), EVS can be used to estimate $\mathrm{P(M_{max}|u_{\delta}})$. $\mathrm{P(M_{max}|u_{\delta}})$ can then be multiplied with $P(\mathrm{u_\delta}|N_{\delta})$ to give us the final estimate of $\mathrm{P(M_{max}|N_{\delta}})$. 

\subsection{Overdensity identification}
\label{subsec:overdensity_ID}

Visual inspection of Figure \ref{fig:EXCELS_overdensity} indicates that the three galaxies of interest reside in overdensities. PRIMER-EXCELS-109760/117560 are close in mass and reside in the same overdensity. ZF-UDS-7329 is significantly more massive than other galaxies belonging to its overdensity.

We fit the size of each overdensity by fitting a spheroidal volume. The spheroidal shape is motivated by theoretical work on clusters and protoclusters in large N-body simulations \citep{Lovell2018_protocluster}. We determine the three major axes of each overdensity by considering the minimum volume enclosing ellipsoid (MVEE) for each overdensity. The volume is grown until there is a clear break in the number of galaxies included, which then defines our \textit{fiducial} overdensity regions. The measurements of the overdensities are presented in Table \ref{tab:overdensity_size}, and visually depicted in Figure \ref{fig:EXCELS_overdensity}. 

However, determining the volume of an overdensity is notoriously tricky. No matter the method, we will always need to make a subjective choice on where to draw a boundary in a continuous density field. Therefore, we also include the measured radii for an \textit{extended overdensity region} where we ignore the first ``gap'' between the central, fiducial overdensity and its surroundings, and grow the overdensity until we again encounter a sharp drop in galaxy number count. These extended overdensities are visually depicted in the bottom row of Figure \ref{fig:EXCELS_overdensity}, and their measurements are likewise included in the bottom section of Table \ref{tab:overdensity_size}. The extended volumes are typically an order of magnitude higher than the fiducial volumes, but, as will be shown in \S \ref{sec:results}, the volume uncertainty has limited influence on our results.

\begin{table}[h!]
    \caption{The fiducial and extended 3D sizes of the two overdensities analyzed in this work. For each overdensity, we also calculate the total volume, as well as the total possible number of such volumes that would fit into the total survey volume at $3<z<5$ ($\mathrm{N_{\delta}}$).}
    \hskip-1.5cm  
    \begin{tabular}{cccccc}
        \multicolumn{6}{c}{\textbf{Fiducial Volumes}} \\
        \hline
        $\langle z \rangle$ & $r_{\mathrm{z}}$ & $r_{\mathrm{DEC}}$ [''] & $r_{\mathrm{RA}}$ [''] & Volume [pMpc$^3$] & $\mathrm{N_{\delta}}$ \\
        \hline
        4.622 & 0.006 & 6 & 130.6 & 0.09 & $3.9 \times 10^5$ \\
        3.197 & 0.073 & 60 & 374 & 88 & 385 \\
        \multicolumn{6}{c}{\textbf{Extended Volumes}} \\
        \hline
        $\langle z \rangle$ & $r_{\mathrm{z}}$ & $r_{\mathrm{DEC}}$ [''] & $r_{\mathrm{RA}}$ [''] & Volume [pMpc$^3$] & $\mathrm{N_{\delta}}$ \\
        \hline
        4.624 & 0.014 & 17 & 592 & 2.7 & $1.3 \times 10^4$ \\
        3.231 & 0.136 & 178 & 741 & 961 & 35 \\
        \hline
    \end{tabular}
    \label{tab:overdensity_size}
\end{table}

Table \ref{tab:overdensity_size} also lists the total number of such subvolumes ($\mathrm{N_{\delta}}$) which would fit inside the EXCELS survey volume from $3<z<5$ ($V_{\mathrm{EXCELS}}=3.4 \cdot 10^4 \mathrm{pMpc}^3$). The volume is calculated as the volume of the photometric catalogue searched for spectroscopic follow-up and is therefore known almost perfectly. $\mathrm{N_{\delta}}$ is central to estimating how extreme the most extreme overdensity of a given volume is.

\subsection{Estimating the percentile of the most extreme overdensity}
\label{subsec:sigma_delta_max}

\begin{figure*}
    \centering
    \includegraphics[trim={0.2cm 7.1cm -6.5cm 7cm},clip, width=1.27\linewidth]{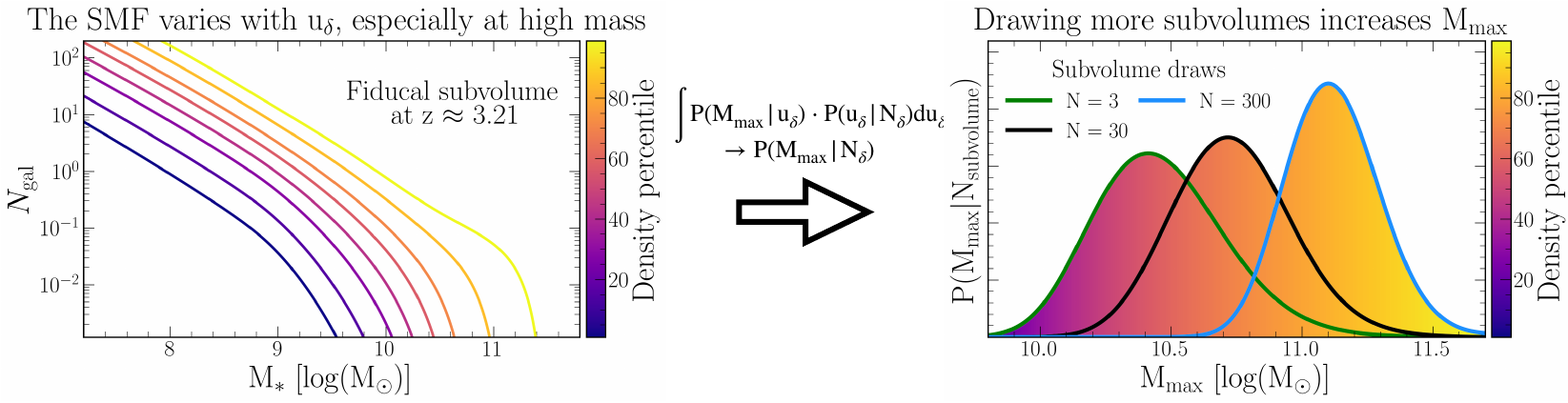}
    \caption{A graphical demonstration of our method for the fiducial subvolume at $z \approx 3.21$. The SMF is varied across different density levels, parametrized by the ``density percentile'', $\mathrm{u}_{\delta}$ (left panel). Each SMF($\mathrm{u}_{\delta}$) has a corresponding distribution for the mass of the most massive galaxy, $\mathrm{P(M_\mathrm{max}|u_{\delta})}$ (not explicitly shown). We can then combine $\mathrm{P(M_\mathrm{max}|u_{\delta})}$ and $\mathrm{P(u_{\delta}|N_{\delta})}$, the distribution of maximum density percentiles given that we observe $N_{\delta}$ such subvolumes. Marginalizing out the laten variable $\mathrm{u}_{\delta}$ then gives us $\mathrm{P(M_\mathrm{max}|N_{\delta})}$, the distribution for the mass of the most massive galaxy given that we observe $N_{\delta}$ subvolumes. In the right panel, we show examples for how $\mathrm{P(M_\mathrm{max}|N_{\delta})}$ evolves, as the number of observed subvolumes (of fixed volume) increases.} 
    \label{fig:demo_fig}
\end{figure*}

To condition on overdensity, we must first estimate how overdense the subvolumes are. However, this is generally not possible to do directly. Instead, we can make the safe assumption that the subvolumes shown in Figure \ref{fig:EXCELS_overdensity} and Table \ref{tab:overdensity_size} are the most overdense of their kind in the survey volume. We can then estimate what density \textit{percentile}, $\mathrm{u}_{\delta}$, that an overdensity likely belongs to, given that it is at the maximum allowed $\mathrm{u}_{\delta}$ in the survey. Since percentiles are by construction uniformly distributed in the [0,1] interval, this corresponds to estimating the maximum draw, $\mathrm{u}_{\delta, \mathrm{max}}$, from sampling a uniform distribution $\mathrm{N_{\delta}}$ times. We can use the Extreme Value Statistics (EVS) formalism to estimate the distribution of $\mathrm{u}_{\delta, \mathrm{max}}$ given $\mathrm{N_{\delta}}$ \citep{gumbel_statistics_1958}. 

We therefore quickly review the basic EVS procedure for an arbitrary distribution. Consider a sequence of \( N \) random variates \(\{X_i\}\), drawn from a cumulative distribution function (CDF), \( F(x) \). There will be a largest value of the sequence, 

\begin{equation}
    X_{\text{max}} \equiv \sup\{X_1, X_2, \dots, X_N\}.
\end{equation}

\noindent If all variables are independent and identically distributed (IID), the probability that all deviations are less than or equal to some value \( x \) is given by:
\begin{eqnarray}
    \hspace{-20pt} 
    \nonumber \Phi(X_{\text{max}} \leq x; N) &=& F(X_1 \leq x) \cdot F(X_2 \leq x) \cdots F(X_N \leq x)\\
    &=& [F(x)]^N    
\end{eqnarray}

\noindent By differentiating, we find the probability density function (PDF) of the maximum value distribution:
\vspace{7pt}
\begin{equation}
    \label{eq:EVS}
    \Phi(X_{\text{max}} = x; N) = N f(x) [F(x)]^{N-1},
\end{equation}

\noindent where \( f(x) = F'(x) \) is the PDF of the original distribution. Note that Eq. \ref{eq:EVS} holds for any distribution f(x), regardless of whether it is a simple probability distribution or a SMF.

We can now construct the distribution $\mathrm{P(u_{\delta, max}|N_\delta)}$ to determine the expected density percentile of the most extreme overdensity, given that $N_\delta$ similar subvolumes are observed. The following analysis can then justifiably be conditioned on $\mathrm{u}_{\delta}$. 

This calculation assumes all subvolumes to be independent, even though galaxy properties are known to correlate with environment on scales of many Mpc \citep{WuJespersen2023_environment,WuJespersen2024_environment}. As a result, these estimates may not be completely accurate. Nevertheless, in a volume as large as the EXCELS $3<z<5$ survey volume, separations that go well beyond the typical strong correlation scales are easily achieved \citep{Lovell2023_EVS}. 
\vspace{7pt}

\subsection{Modeling the most massive galaxy in an overdensity}
\label{subsec:model}

To estimate $\mathrm{P(M_{max}|N_{\delta}})$, we must now model the number counts of galaxies in a given volume, the stellar mass function (SMF). To construct an overdensity-dependent SMF, we must first adopt a mean SMF. The mean SMF is made by directly linking the mean SMF to the halo mass function (HMF), assuming a one-to-one mapping between galaxy and halo masses. The linking depends on the baryon fraction ($f_b$) \citep{Planck2020}, and the stellar baryon fraction\footnote{SBF, SFE, and $\epsilon_*(z)$ are all used interchangably here.}, $\epsilon_*$:

\begin{equation}
    \label{eq:ms_mh}
    M_*(z) = f_b \cdot \epsilon_*(z) \cdot M_{\mathrm{halo}} .
\end{equation}

We use the observed $\epsilon_*(z)$ of \cite{Finkelstein2015_SBF}, $\epsilon_*(z) = 0.051 + 0.024 \cdot (z - 4)$. This relation was calibrated at $4\leq z \leq7$, and since \cite{Finkelstein2015_SBF} included several galaxies of $\mathrm{M_*}>10^{11} \mathrm{M}_{\odot}$ when fitting their relation, this pre-\textit{JWST} $\epsilon_*(z)$ should be well-suited for the galaxies analyzed here. Parameter uncertainties could be propagated through our analysis, but their effect is subdominant compared to the change induced by conditioning on environment. For the HMF, we use the \cite{Behroozi2013_HMF} model. Some uncertainty is expected in the HMF in the ultra-high-z regime \citep{Yung24_highz}, but for relatively massive halos at $z<10$, the focus of this paper, most HMFs agree closely.
To obtain the mean number density of galaxies in a given mass bin from $M_i$ to $M_j$, we can then integrate the SMF between these two masses. To get the absolute number, this is multiplied by the volume in question.

To model the real universe, we must adopt a distribution for the number of galaxies in a given mass bin. If galaxies did not cluster, the Poisson distribution would be the appropriate choice; however, since galaxies cluster quite strongly, we must adopt a Poisson-like distribution which allows for super-Poissonian variance. This additional variance is usually known as \textbf{cosmic variance}, $\sigma_{\mathrm{CV}}$. $\sigma_{\mathrm{CV}}$ is typically given as a fractional variance \citep{Moster2011}. It has been shown by \cite{Steinhardt2021} and \cite{Jespersen2025_CV_highz} that the gamma distribution is an appropriate model for the distribution of galaxy number counts in the presence of cosmic variance. We therefore adopt this distribution, defined as:

\begin{equation}
    \label{eq:gamma}
    f(x;\mu,\sigma) = f(x;k,\theta) = P(X=x) = \frac{x^{k-1}e^{-\frac{x}{\theta}}}{\theta^k\Gamma(k)},
\end{equation}

\noindent where $\Gamma$ is the gamma function, and $k$ and $\theta$ are shape and scale parameters which independently govern the mean ($\mu = k\theta$) and variance ($\sigma^2 = k\theta^2$) of the distribution. $\mu$ is calculated from the mean SMF, and the variance of the distribution in the presence of cosmic variance is $\sigma^2 = \mu + \sigma_{\mathrm{CV}}^2\mu^2$ \citep{Jespersen2025_CV_highz}. For high-mass, high-z galaxies, we generally have $\sigma_{\mathrm{CV}}^2\mu^2>>\mu$, reflecting their high clustering amplitude.

We can now construct SMFs \textit{dependent on density percentile}. The spread in the distribution of galaxy number counts mainly originates from variations in the underlying density field, as evidenced by the fact that galaxy number counts in all mass bins are tightly correlated \citep{Thomas2023_FLARES_overdensity, Jespersen2025_CV_highz}. The variation in number counts across different environments is thus effectively modeled by the variation in the gamma distribution. We build SMFs dependent on density percentile by considering the corresponding percentile of the gamma distribution for each mass bin. In short, for each mass bin, we take $\mathrm{N_{gal}}$ where the CDF of the gamma distribution for that mass bin is equal to $\mathrm{u}_\delta$. In this way, the density percentiles, $\mathrm{u}_{\delta}$, define a self-consistent set of SMFs which vary only with $\mathrm{u}_{\delta}$. This is shown in the left panel of Figure \ref{fig:demo_fig}.

Because $\sigma_{\mathrm{CV}}$ is higher for higher masses, the gamma distribution has a higher variance for higher masses. This then implies that the SMF for high $\mathrm{u}_{\delta}$ boosts high-mass galaxy formation, rendering a typical double Schechter function shape with a pronounced ``knee'', as can be seen in Figure \ref{fig:demo_fig}. Similarly, high-mass galaxy formation is excessively suppressed for low $\mathrm{u}_{\delta}$, as is expected from simulations \citep{lovell_first_2021, YuePan2025_void_galaxies}.

For each SMF($\mathrm{u}_{\delta}$), we can now perform the well-established EVS technique introduced by \cite{Lovell2023_EVS} to calculate $P(M_\mathrm{max}|{\mathrm{u}_{\delta}})$. Since $\mathrm{u}_{\delta}$ is not known, we instead calculate $P(\mathrm{u}_{\delta,\mathrm{max}}|\mathrm{N}_{\delta})$ as described in \S \ref{subsec:sigma_delta_max}. We only need to know $\mathrm{N}_{\delta}$, the number of subvolumes with the same volume as the overdensity of interest which fit inside the full survey volume. Using these two distributions, we can then construct the more relevant distribution:
\begin{eqnarray}
&& P(M_{\mathrm{max}}\,|\,\mathrm{N}_{\delta}) =   \\
 & &~~~~~~~~~~ \int_0^1 
P(M_\mathrm{max}\,|\,\mathrm{u}_{\delta,\mathrm{max}})\, 
 \cdot P(\mathrm{u}_{\delta,\mathrm{max}}\,|\,\mathrm{N}_{\delta})\, 
d\mathrm{u}_{\delta,\mathrm{max}} \nonumber
\end{eqnarray}

This distribution is shown in the right panel of Figure \ref{fig:demo_fig}. To construct the evolution with redshift, we simply repeat the calculation for different $z$. 

The code implementing this extended version of EVS can be found either via Zenodo \citep{Jespersen2025_EVS_code}, or via \href{https://github.com/astrockragh/evs_clustering}{GitHub}.

\subsubsection{Calibrating the cosmic variance}
\label{subsec:calibrate_CV}

Massive quiescent galaxies form preferentially in regions of high density and are thus excessively biased. Their cosmic variance is therefore very high, since cosmic variance for galaxies is intrinsically connected to galaxy bias ($b$) through

\begin{equation}
    \label{eq:sigma_CV_bias}
    \sigma_{\mathrm{CV}} = \sigma_{\mathrm{CV,gal}} = b\cdot\sigma_{\mathrm{CV,DM}}
\end{equation}

\noindent and bias is known to scale with mass and color \citep{Beisbart2000_quiescent_bias, Hu2003_bias_with_mass, Wang2010_quiescent_bias}.
There are many ways of calculating $\sigma_{\mathrm{CV}}$, including using empirically calibrated calculators \citep{trenti_cosmic_2008, Moster2011}, and calibrating to simulations \citep{Bhowmick2020, Jespersen2025_CV_highz}. While empirically calibrated calculators are generally preferable, there are currently no cosmic variance calculators calibrated in the applicable range of redshift and mass. Outside of their calibration range (in either redshift or stellar mass), they are usually not reliable. Therefore, we follow \cite{Jespersen2025_CV_highz} and calibrate the cosmic variance using the publicly available catalogs from the \texttt{UniverseMachine}\footnote{\url{https://halos.as.arizona.edu/UniverseMachine/DR1/JWST_Lightcones/}} simulation suite \citep{Behroozi2019_UniverseMachine}. \texttt{UniverseMachine} is tuned to reproduce observed galaxy clustering, which is the defining feature for cosmic variance.

We use a total of 32 \texttt{UniverseMachine} lightcones, consisting of eight realizations with different initial conditions from each of four different survey designs, \texttt{COSMOS} (17'x41'), \texttt{UDS} (36'x35'), \texttt{GOODS-N} (69'x32') and \texttt{GOODS-S} (45'x41'). We split each lightcone into sample volumes paralleling the volumes analyzed here and sample the number count distribution. We then calibrate $\sigma_{\mathrm{CV}}$ directly to the sampled number count distributions. However, since $\sigma_{\mathrm{CV}}$ is very high for the small fields and massive galaxies considered here, we must apply the corrections identified by \cite{Jespersen2025_CV_highz} when interpolating relations between the sampled redshifts/masses, such that each sampled $\sigma_{\mathrm{CV}}$ is weighted by the uncertainty on the estimate of the cosmic variance, $\sigma_{\sigma_{\mathrm{CV}}}= \sqrt{\frac{ \sigma^2_{\mathrm{CV}}+1}{N}}$.

\section{Results}
\label{sec:results}

\begin{figure*}
    \centering
    \includegraphics[trim={0cm 0cm 0cm 0cm},clip, width=1\linewidth]{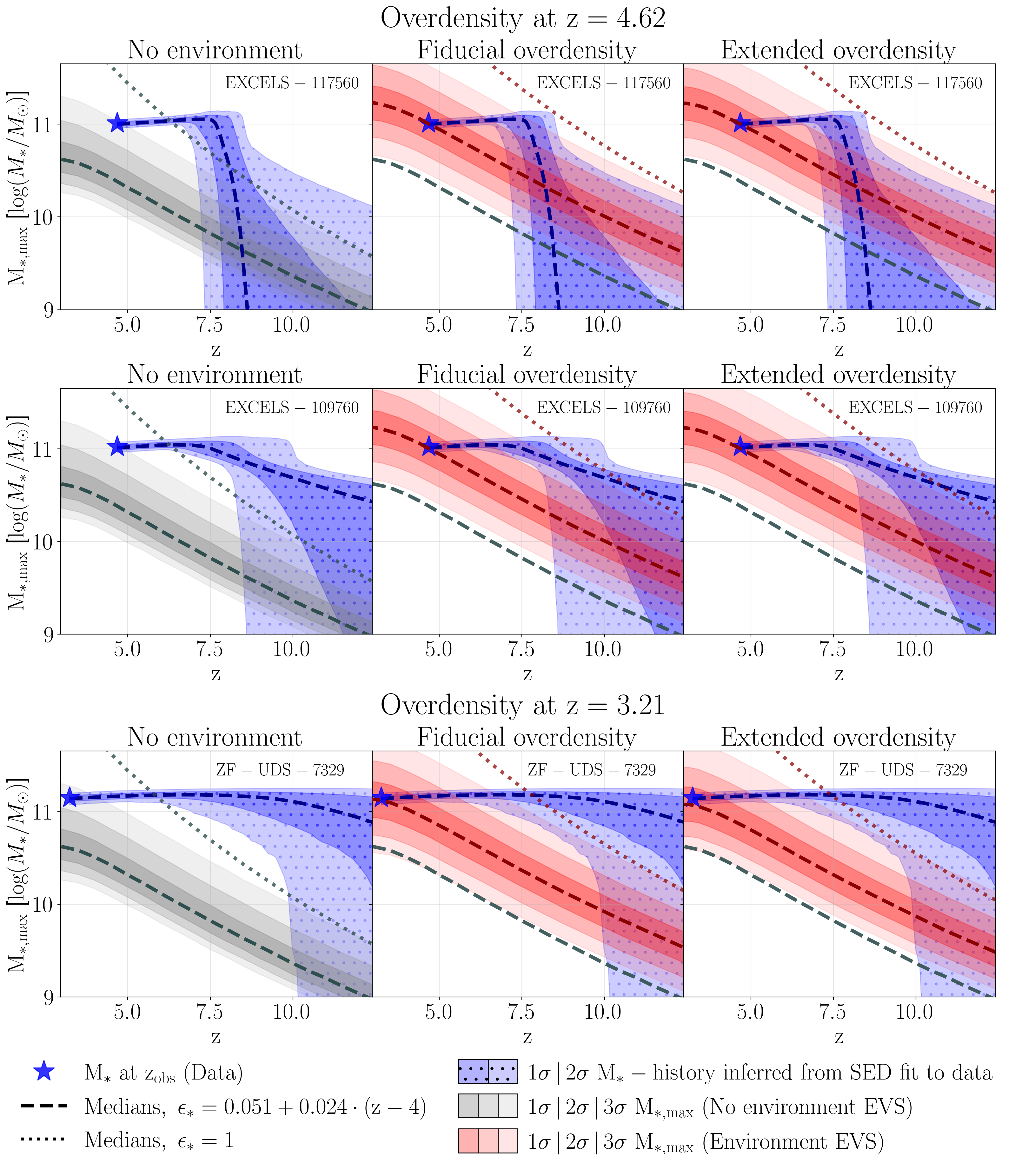}
    \caption{Inferred stellar mass histories of the EXCELS ultra-massive quiescent galaxies (blue, dotted) along with different potential theoretical models for their masses (red and grey). Modelling extreme galaxies in the context of their environments (red) raise their expected masses relative to the standard EVS framework (grey), which does not take environment into account. In general, none of the galaxies seem extremely overmassive at the redshift at which they are observed (blue stars), but in the standard, unclustered, EVS framework (left-most column), these galaxies would be overmassive at a level above $5\sigma$ at $6<z<10$. Some tensions persist even when setting $\epsilon_*=1$. However, when taking into account the overdensities these galaxies inhabit (right columns), we observe only mild tensions of $\approx2\sigma$. As is shown in the right-most column, our results are robust to order-of-magnitude uncertainties in the overdensity volumes.} 
    \label{fig:EXCELS_highz_mostmassive}
\end{figure*}

Here we show results for the two overdensities which contain the ultra-massive quiescent EXCELS galaxies. 
We compare the stellar mass histories inferred from SED fitting\footnote{We use the \texttt{BAGPIPES} code for fitting the EXCELS spectra \citep{Carnall2018_bagpipes}.} with the redshift evolution of the most massive galaxy predicted by EVS, with and without conditioning on environment.

\subsection{PRIMER-EXCELS-109760/117560}
\label{subsec:results_109760_117560}
The most distant overdensity, at $z = 4.62$, contains PRIMER-EXCELS-109760 and PRIMER-EXCELS-117560. The fiducial volume contains 8 total galaxies and is highly concentrated. As shown in Table \ref{tab:overdensity_size}, $3.9\cdot10^5$ similar subvolumes could be contained in the survey volume. The mass-redshift contours for the galaxies in this subvolume are shown in the first two rows of Figure \ref{fig:EXCELS_highz_mostmassive}. The inferred stellar mass histories are plotted alongside the modelled galaxy masses as a function of redshift. The first column shows the expected mass using standard EVS, whereas the second column shows the expected mass when we condition on the galaxies existing in the fiducial overdensity.
We can also postulate that the original overdensity volume is underestimated and instead work with the extended volume from Table \ref{tab:overdensity_size}, which contains 18 galaxies. The results for the extended volume are shown in the third column of Figure \ref{fig:EXCELS_highz_mostmassive}. There is very little difference between the results from the fiducial and extended volumes. We thus conclude that our results are robust to the uncertainty in volume. 

Figure \ref{fig:EXCELS_highz_mostmassive} clearly shows that both galaxies at $z \approx 4.62$ have natural masses consistent with a pre-\textit{JWST} SFE, \textit{if} one considers the overdensity they inhabit. To directly compare with the standard EVS approach, we can consider the maximum tension between model and data for each galaxy. For standard EVS, we get $\sigma_{\mathrm{max}}$ of 5.7 and 5.0, for 109760 and 117560, respectively. By conditioning on overdensity, we get $\sigma_{\mathrm{max}}$ of 2.4 and 1.7. We have thus moved from strong tension to only a mild suggestion of tension. Notably, this alignment does not require an elevated SFE relative to the pre-\textit{JWST} estimates by \cite{Finkelstein2015_SBF}.

This initially ignores any potential scatter in the $M_\mathrm{h}-M_*$ relation. While this is on the order of 0.3 dex for the broad galaxy population \citep{Jespersen2022, chuang2024_mangrove}, it is significantly smaller for massive quiescent galaxies ($\sigma_{M_\mathrm{h}-M_*}=0.1$ dex, \cite{Chuang2022}). Including a scatter of $\sigma_{M_\mathrm{h}-M_*}=0.1$ dex, the tensions drop to 5.2$\sigma$ and 5.1$\sigma$ for the standard EVS, and 2.2$\sigma$ and 1.5$\sigma$ for the model conditioned on overdensity. If we are willing to raise the stellar baryon fraction to 100\%, we get 1.41$\sigma$ and 0.49$\sigma$ for the standard EVS, and 0.0 $\sigma$ and -1.1$\sigma$ for the model conditioned on overdensity. An expanded table of tensions for different models can be found in Appendix \ref{appsec:tension_table}.

\subsection{ZF-UDS-7329}
\label{subsec:results_55410}
The overdensity at $z=3.21$ contains ZF-UDS-7329, with the fiducial volume consisting of 63 total galaxies. Since there are many more galaxies in this overdensity, and the incompleteness is much lower, the radii in all directions are likely much more robust. The probability contours for the mass of the most massive galaxy are shown in the bottom row of Figure \ref{fig:EXCELS_highz_mostmassive}. The possible extended overdensity volume containing ZF-UDS-7329 consists of 190 galaxies and is an order of magnitude larger in volume. However, just as with the high-z overdensity, this change makes only a small difference, and the differences between the contours in the last two panels of the last row of Figure \ref{fig:EXCELS_highz_mostmassive} are barely distinguishable. We thus conclude that our results are robust against any realistic uncertainty in volume.

We can again calculate the maximum tension between our model and the observed masses. With standard, unclustered EVS, ZF-UDS-7329 presents serious tension with a $\sigma_{\mathrm{max}}$ of 5.7, but when conditioning on overdensity, this drops to 3.2. A 3$\sigma$ discrepancy is noteworthy, but does not constitute a significant theoretical challenge.

Including a scatter of $\sigma_{M_h-M_*}=0.1$ dex, the tension drops to 5.2$\sigma$ for the standard EVS, and 3.0$\sigma$ for the model conditioned on overdensity. If we are willing to raise the stellar baryon fraction to 100\%, we get 2.42$\sigma$ for the standard EVS, and 0.7$\sigma$ for the model conditioned on overdensity. Thus, even when making such extreme modifications as having 100\% star formation efficiency, the normal EVS approach still yields mild tension, while the overdensity model yields zero tension.

\section{Discussion}
\label{sec:discussion}

The results of this paper show that the extreme masses of high redshift, quiescent galaxies can be explained without resorting to extreme star formation efficiencies, \textit{if} one considers the environments of these galaxies. Specifically, one must be willing to condition on a given extreme galaxy residing in the most extreme overdensity in its field. This clearly demonstrates that no galaxy should be considered as an island, each instead existing in a fabric of relations \citep{lyotard1984postmodern}. However, there are still some caveats to be aware of.

First, in addressing what might appear to be a significant uncertainty in our model due to the difficulty in estimating overdensity volumes, our analysis (see Figure \ref{fig:EXCELS_highz_mostmassive}) demonstrates that even an order of magnitude error in the volume leads to only small changes in the mass-redshift contours. The robustness to volume misestimation further reinforces the reliability of our overall conclusions.

Another key assumption of our model is that the analyzed volumes are the most overdense of their kind in the EXCELS survey volume. We have verified that with currently available spectroscopic data, the fiducial overdensity volumes are the most dense of their kind in the full survey volume. We test this by randomly placing randomly rotated ellipsoids with the same shape as our fiducial ellipsoid across the survey volume. For every sampled ellipsoid, the number of galaxies within the ellipsoid is counted. This sampling is then repeated $10^{6.5}$ times, roughly ten times more than $\mathrm{N_{\delta}}$ for the fiducial subvolume at z=4.622. We observe that no randomly sampled volume ever contains as many galaxies as our fiducial volumes. Figures showing the distributions of sampled number counts can be found in Appendix \ref{appsec:sample_subvolumes}. Our fiducial volumes, the most dense by number counts, are thus likely to be the most dense by mass, due to the tight relationship between cluster richness, mass, and the mass of the most massive cluster galaxy \citep{Yee2003_mass_richness, Andreon2010_mass_richness}.

Furthermore, the SFEs that must be assumed to not have any significant tension are still higher than low redshift analogs, which typically have SFEs on the order of a few percent \citep{behroozi_comprehensive_2010, Shuntov_2022_Mh_Ms, Shuntov_2025_Mh_Ms}. Therefore, while the need to increase SFE with redshift is mitigated, it is not fully resolved \citep{steinhardt_impossibly_2016}, although the required increase has been lowered by almost an order of magnitude.

A key uncertainty in our model is the calibration of the cosmic variance. Although calibrating to \texttt{UniverseMachine} is currently the best available option, it still introduces a simulation-dependent cosmic variance. Luckily, both upcoming and ongoing \textit{JWST} surveys like PANORAMIC will focus on clustering and cosmic variance \citep{Williams2025_panoramic}, while the first high-z galaxy bias measurements have just been made using COSMOS-Web by \cite{Paquerau2025_cosmos_clustering}. The mass bins used by \cite{Paquerau2025_cosmos_clustering} are unfortunately too incomplete for our work, but it shows that JWST-relevant galaxy bias calibrations are becoming possible. Naturally, clustering amplitudes based on wide-band photometry are very uncertain, but upcoming spectroscopic and medium band surveys such as CAPERS \citep{CAPERS_JWSTprop} and MINERVA\footnote{\href{https://www.stsci.edu/jwst-program-info/download/jwst/pdf/7814/}{www.stsci.edu/jwst-program-info/download/jwst/pdf/7814/}} should be able to offer significantly better constraints.

In addition, significant systematic uncertainties persist in the SED-fitting process; various assumptions can lead to considerable changes in both the inferred masses and other galaxy properties \citep{ConroyGunn2009a_SSPI_uncertain_parameters, ConroyGunn2010_SSPII_comparing_observations_to_models,Steinhardt2023_IMF_highz, Iyer2025_review, Jespersen2025_IR_optical}. Fortunately, quiescent galaxies tend to be less affected by these systematics compared to their star-forming counterparts.

We also treat each galaxy as the most massive in our survey at a given redshift, but this cannot be the case for both PRIMER-EXCELS-109760 and PRIMER-EXCELS-117560. One could thus imagine splitting the subvolume in two to assign half of the subvolume to each galaxy. Doing so has minimal impact on the resulting probabilities, only lowering the contours by a few percent. This reflects that the mass of the second most massive galaxy in a given volume should be quite close to the mass of the most massive galaxy, since neighboring mass bins are very tightly correlated \citep{Thomas2023_FLARES_overdensity, Jespersen2025_CV_highz}. 

Due to the extremely compact nature of these overdensities, it is also likely that our galaxies have experienced significant merging in the past, which could have enhanced the stellar masses \citep{Ito2025_cosmic_vine_mergers}. Our analysis assumes that all stars have formed \textit{in situ}, since this is the simplest and most conservative assumption. However, this should not result in any major error, since \cite{Baker2025_notmergers} have shown that most massive quiescent \textit{JWST} galaxies formed the vast majority of their stars \textit{in situ}, similarly to the ultra-massive quiescent galaxy of \cite{deGraaff2025}.

A future improvement to the model could be to go further in the conditioning scheme, conditioning not only on overdensity, but also on color and size. Constructing SMFs conditional on these quantities would have to be largely empirical. Although these empirical relations exist \citep{Weaver2023_COSMOS_SMF, Ito2024_size_stellar_mass}, their interpretations are not as clean as the relationship between overdensities and number counts, as the drivers of quiescence and morphological evolution are still unclear \citep{HuertasCompany2024_JWST_morphology, Valentino2025_gas_outflows_highz_quiescent}.
\section{Conclusion}
\label{sec:conclusion}
We conclude that the masses of ultra-massive quiescent galaxies at $3<z<5$ do not represent a significant tension with simple theoretical models of galaxy formation, \textit{if} one is willing to condition the analysis of any given galaxy on its environment. We have shown that the extreme galaxies presented by \cite{Carnall2024_EXCELS} and \cite{Glazebrook2024_highmass_quiescent_galaxy} reside in extreme overdensities, which allows us to perform this conditioning step. When conditioning on environment, the SFE required to explain these galaxies drop from 100\% to $\approx 10\%$ in the high-tension regime of $6<z<10$, as per the pre-\textit{JWST} SFE-redshift relation provided by \cite{finkelstein_case_2015}.
This highlights the importance of considering the relation between galaxies and their environments.

\section{Acknowledgements}

The authors wish to thank the organizers of the 40th IAP Symposium for providing the possibility for the authors to meet. The authors furthermore wish to thank Adrian E. Bayer, David N. Spergel, Doug Rennehan, Romain Teyssier, Andrew K. Saydjari, and Marielle Côté-Gendreau for discussions and comments, which greatly improved the paper. ACC acknowledges support from a UKRI Frontier Research Guarantee Grant (PI Carnall; grant reference EP/Y037065/1). The code implementing our extended version of EVS can be found either via Zenodo \citep{Jespersen2025_EVS_code}, or via \href{https://github.com/astrockragh/evs_clustering}{GitHub}, both of which contain a demonstration notebook on how to use the software.

\textit{Software:} \texttt{matplotlib} \citep{matplotlib}, \texttt{pandas} \citep{pandas}, \texttt{jupyter} \citep{jupyter}, \texttt{numpy} \citep{numpy}, \texttt{AstroPy} \citep{2022_Astropy}, \texttt{HMFcalc} \citep{Murray2013_hmfcalc}, \texttt{scipy} \citep{scipy}. 

\bibliographystyle{aasjournal}
\bibliography{main}



\appendix

\section{Tensions for different models}
\label{appsec:tension_table}

\begin{deluxetable}{cccccccccc}[h!]
\tablecaption{Maximum tension between a given model and each observed galaxy (in Gaussian units). The stacked row assumes all measurements are independent, representing an overestimate of the total tension. To achieve the same tensions in standard EVS and in the overdensity Fiducial Model, the EVS SBF would need to be elevated to 80\%, an order of magnitude above that of the fiducial model. \label{tab:many_models}}
\tablehead{
\colhead{} & 
\multicolumn{4}{c}{EVS Conditioned on Overdensity} & 
\multicolumn{5}{c}{Standard EVS} \\
\cline{2-5} \cline{7-10}
\colhead{Galaxy ID} &
\colhead{Fiducial} & 
\colhead{$\sigma_{\mathrm{M_H-M_*}}=0.1$} & 
\colhead{$\sigma_{\mathrm{M_H-M_*}}=0.3$} & 
\colhead{SBF=1} & \colhead{    $~~$    } &
\colhead{Fiducial} & 
\colhead{$\sigma_{\mathrm{M_H-M_*}}=0.1$} & 
\colhead{$\sigma_{\mathrm{M_H-M_*}}=0.3$} & 
\colhead{SBF=1}
}
\startdata
ZF-UDS-7329    & 3.17 & 3.00 & 2.27 &  0.69 &   & 5.72 & 5.23 & 3.54 & 2.42 \\
EXCELS-109760  & 2.39 & 2.16 & 1.38 &  0.01 &    & 5.74 & 5.09 & 3.10 & 1.41 \\
EXCELS-117560  & 1.70 & 1.52 & 1.02 & -1.10 &    & 5.01 & 4.41 & 2.67 & 0.49 \\
Stacked        & 4.32 & 4.00 & 2.85 &  0.69 &     & 9.53 & 8.53 & 5.41 & 2.84 \\
\enddata
\end{deluxetable}

We can consider several different models to explore the tensions with our observed galaxies. The most interesting statistic is the \textbf{maximum tension} between any observed galaxy and corresponding model. These are shown in Table \ref{tab:many_models}, where we include our fiducial model, models with 100\% star formation efficiency, and models with different levels of scatter between halo and stellar masses. 

Interestingly, the fiducial model \textit{conditioned on overdensity} produces tensions comparable to the standard EVS model with SBF = 1. In fact, using SBF = 0.8 with standard EVS gives almost the same results as the fiducial SBF model combined with conditioning on overdensity. Conditioning on overdensity therefore affords us roughly an order of magnitude in star formation efficiency. 

\section{Sampling similar subvolumes}
\label{appsec:sample_subvolumes}

\begin{figure}[h!]
    \centering
    \includegraphics[width=0.8\linewidth]{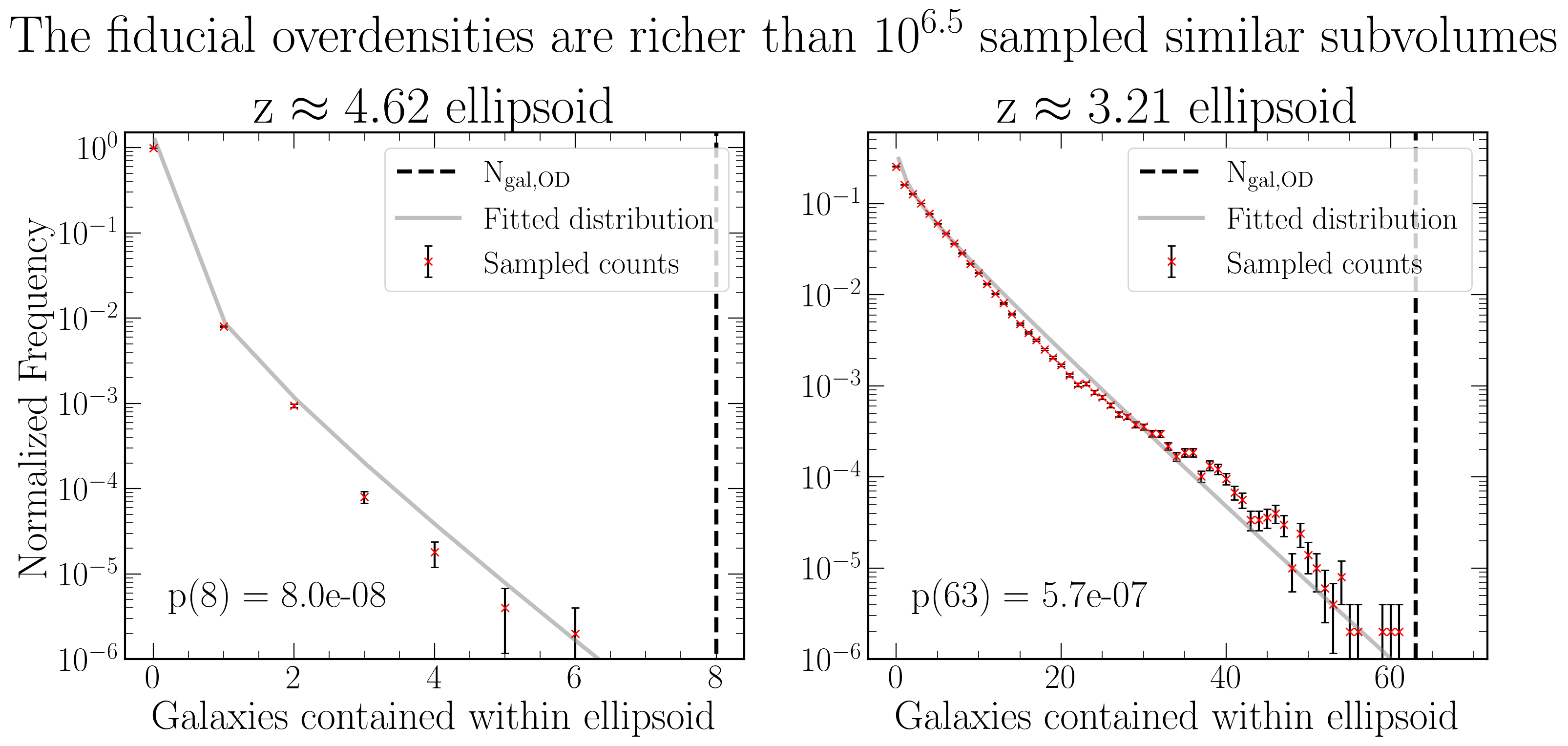}
    \caption{Distributions of galaxy counts within $10^{6.5}$ randomly sampled ellipsoids (red points with error bars) at $z \approx 4.62$ (left) and $z \approx 3.21$ (right). Dashed vertical lines show the number of galaxies in the fiducial overdense volumes, which are unmatched in the sampling. Grey curves show gamma function fits to the count distributions, with evaluated probabilities at the fiducial counts annotated. For the $z \approx 3.21$ subvolume, the samples that approach the fiducial number count are all very close to the fiducial subvolume.}
    \label{fig:random_sample_subvolumes}
\end{figure}

A key assumption of our analysis is that the fiducial overdense regions we study represent the most extreme overdensities of their kind within the EXCELS survey volume. To quantitatively validate this, we conduct a Monte Carlo test of overdensity rarity using the currently available spectroscopic galaxy catalog.

We define ellipsoids with identical size as our fiducial structures at $z \approx 4.62$ and $z \approx 3.21$. We then randomly sample the survey volume with these ellipsoids, applying random positions and orientations, and count the number of galaxies enclosed within each sampled ellipsoid. This sampling process is repeated $10^{6.5}$ times for each fiducial subvolume.

In each case, we find that no randomly sampled volume contains as many galaxies as the fiducial overdensities. This result confirms that the fiducial volumes are the most extreme in terms of galaxy number counts. The galaxy counts from the sampling are shown in Figure~\ref{fig:random_sample_subvolumes}, along with vertical dashed lines indicating the richness of the fiducial overdensities. We also fit gamma distributions to the Monte Carlo results and evaluate the probability of obtaining the observed counts. 

For the $z \approx 4.62$ ellipsoid, the fiducial volume contains 8 galaxies, while the fitted probability density at this value is $p(8) \approx 8.0 \times 10^{-8}$. Since this overdensity is extremely compact, it is natural that it is very hard to reach the richness of the fiducial subvolume. For the $z \approx 3.21$ volume, the fiducial count is 63 galaxies, with a fitted probability $p(63) \approx 5.7 \times 10^{-7}$. Given that this subvolume is significantly larger, it is unsurprising that samples with this volume almost reach the richness of the fiducial subvolume. For the $z \approx 3.21$ subvolume, the samples that approach the fiducial number count are all very close to the fiducial subvolume.

In both cases, the overdensities are thus extremely rare. Given the tight correlation between richness and mass \citep{Yee2003_mass_richness, Andreon2010_mass_richness}, we conclude that our fiducial regions are very likely to also be the most overdense by mass.

\end{document}